# Precision Empirical Mass Formulae for Baryon Octet and Decuplet

Sung La
Physics Department, William Paterson University
Wayne, New Jersey 07470

Abstract

Empirical mass formulae for the baryon octet and decuplet are presented. These formulae are functions of one integer variable and charge state of the baryon. With an exception of $\Lambda(1116)$, the formulae generate masses within 0.1% of the observed masses. The formulae also generate the same electromagnetic mass splittings predicted by SU(6) model. Spin ½ octet resonances and its relation to the octet mass formula is described.

## I. Mass Formulae

In quark models, the baryon octet and decuplet are three-quark states *(u,d,s)* and there are number of very different model calculations for the baryon masses. For some representative calculations, refer to Ref.[1-4]. Agreements with the observed masses are generally quite good at the level of 1% or better.

In this paper, it shown that it is possible to construct entirely empirical mass formulae, independent of any specific model as functions of one integer variable assigned to each member of the baryon sets and charge state of the baryon. The formulae generate masses that are, with an exception of $\Lambda(1116)$, in agreement with the observed masses up to five digits.

As mentioned above, we assign an integer number $i$ which may be called *"mass number"* to each member of the baryon sets as shown in the following Table I. The $j$ numbers in the table are the electric charges (i.e., for the proton, $j=+1$; for the neutron, $j=0$, etc.). In the table, the baryon sets are grouped together in a most compact form such a way that $i$ number runs from 0 to 3 for the octet members and from 1 to 4 for the ducuplet members.

| j \ i | 0 | 1 | 2 | 3 | 4 |
|---|---|---|---|---|---|
| +2 |  | $\Delta^{++}$ |  |  |  |
| +1 | $p$ | $\Delta^+$ | $\Sigma^+, \Sigma^{*+}$ |  |  |
| 0 | $n$ | $\Delta^0, \Lambda$ | $\Sigma^0, \Sigma^{*0}$ | $\Xi^0, \Xi^{*0}$ |  |
| -1 |  | $\Delta^-$ | $\Sigma^-, \Sigma^{*-}$ | $\Xi^-, \Xi^{*-}$ | $\Omega^-$ |

TABLE I. Assignment of the *"mass number" i. j* numbers are the electric charges (i.e. for the proton, $j=+1$; for the neutron, $j=0$, etc.).

For convenience, in the following discussion we let the symbols of the baryons stand for their masses. Using the *i* and *j* numbers, the mass of each member of the baryon sets can now be represented as *M(i, j)* for the octet and *M\*(i, j)* for the decuplet members (e.g., *M(0, 1) = p; M\*(3,-1)=* $\Xi^{*-}$, etc.).

For the baryon octet, the mass difference between $M(i, j)$ and $M(i\pm 1, j)$ is around 120 MeV and that between $M(i, j)$ and $M(i, j\pm 1)$ is in the ragne of 1-7 $(MeV)$. By careful observation on how these differences vary from one baryon to the next in an accurate manner, we find that mass of the octet baryons are dependent upon $i$ and $j$ numbers in the following formula that involves three constants, $A$, $B$, and $C$,

$$M(i, j) = p + A + Bi - \frac{Ci(i-1)(i-2)}{6} + \frac{(C+e)j(j-1)}{4} - \frac{Ci(j-1)}{2}, \qquad (1)$$

where  $C = 2(n-p) + e = 3.09765892\ (MeV)$,
  $A = C - (1/2)(C + e)\delta_i - (1/2)(C - e)\delta_i\delta_{j-1}$.
  $\delta_i$ and $\delta_{j-1}$ are the Dirac delta functions (i.e., $\delta_k = 1$ if $k = 0$; $\delta_k = 0$ if $k \neq 0$),
  $B$ is a constant to be determined by an input parameter.

As shown above, the constants $A$ and $C$ are related to the masses of $p$, $n$ and in addition to the electron mass $e$. In order to fix the value of $B$, we can choose as an input parameter any one of the octet masses other than $p$ and $n$ which are already incorporated in $C$. It seems that the octet baryon mass whose experimental value has not been changed over more than a decade is $\Sigma^+(1189.370)$ [5]. So we use this mass for $B$. Noting that $i$ and $j$ numbers for $\Sigma^+$ are 2 and 1, respectively according to the Table I, it is simple to obtain the following using (1),

  $\Sigma^+ = M(2, 1) = p + C + 2B$,
  $B = (1/2)(\Sigma^+ - p - C) = 124.00015\ (MeV)$.

Thus, three constants in the expression of $M(i, j)$, namely $A$, $B$ and $C$ are related to three input baryon masses, $p$, $n$ and $\Sigma^+$. Numerical values given above for $B$ and $C$ are obtained using the observed masses in PDG [5]. Table II shows the comparison of the calculated massess of $M(i, j)$ with the observed masses PDG[5-7]. With an exception of $\Lambda(1116)$, agreement is excellent.

TABLE II
Baryon Octet Masses

| Observed Masses (MeV) | M(i, j) Masses |
|---|---|
| $p$  938.27203(a) | $M(0,1)$= 938.27203 (input) |
| $n$  939.56536(a) | $M(0,0)$= 939.56536 (input) |
| $\Lambda$  1115.683 ± 0.006(a) | $M(1,0)$=1066.918 |
| $\Sigma^+$  1189.370 ± 0.07(a) | $M(2,1)$=1189.37    (input) |
| $\Sigma^0$  1192.642 ± 0.024(a) | |
| 1192.460 ± 0.08(b) | $M(2,0)$=1192.467 |
| $\Sigma^-$  1197.449 ± 0.03(a) | |
| 1197.340 ± 0.05(b) | $M(2,-1)$=1197.369 |
| $\Xi^0$  1314.830 ± 0.2(a) | |
| 1314.900 ± 0.6(c) | $M(3,0)$=1314.918 |
| $\Xi^-$  1321.310 ± 0.13(a) | $M(3,-1)$=1321.369 |

(a) Particle Data Group, Physics Lett. **592B** (2004).
(b) Particle Data Group, Physics Lett. **204B** (1988).
(c) Particle Data Group, European Physical Journal,**C3** (1998).

For the decuplet members, detailed observation is made on how a mass of one octet baryon changes into that of corresponding decuplet baryon at a given set of $i$ and $j$ numbers instead of one ducuplet member to the next as done for the octet case. The mass formula thus obtained for the decuplet baryons is given as the following with two new constants $A^*$ and $B^*$,

$$M^*(i,j) = p + A^* + B^* i - \frac{(C + \frac{e}{3})i(i-1)(i-2)}{4} + \frac{(C+e)j(j-1)}{4} - \frac{(C-e)(j-1)}{2}, \quad (2)$$

where $A^* = A + D^*(1 - \delta_i \delta_{j-1})$, the $A$ in this expression and $C$ in the above expression are the same as given in the octet case. $B^*$ and $D^*$ are the new constants to be determined by two new input masses. One noticeable difference in the expression of $M^*(i, j)$ from that of $M(i, j)$ is that the coupling between $i$ and $j$ shown in the last term of (1) is not involved. In order to fix the values of $B^*$ and $D^*$, we once again can choose any two of the decuplet baryon masses. It appears that $\Sigma^{*+}$ and $\Xi^{*0}$ have slightly less uncertainty in their observed masses [5] than others, so we use those for the $B^*$ and $D^*$. We then obtain readily the following using (2),

$\Sigma^{*+} = M^*(2,1) = p + C + D^* + 2B^*,$
$\Xi^{*0} = M^*(3,0) = p + D^* + 3B^* - e.$

From the above two relations, $B^*$ and $D^*$ are given as,

$B^* = \Xi^{*0} - \Sigma^{*+} + C + e = 152.608657 \ (MeV).$
$D^* = 3\Sigma^{*+} - 2\Xi^{*0} - p - 3C - 2e = 136.212995 \ (MeV).$

Table III shows the comparison of $M^*(i, j)$ values with the observed masses of the baryon decuplet. Agreement is about, with the exception of $\Delta^+$ and $\Delta^0$ as good as the octet baryons. Why the lesser agreement involving $\Delta^+$ and $\Delta^0$ is not clear, but, in the Table III it may be noteworthy that for $\Delta^+$, the uncertainty in its observed mass is either not given or roughly seven times larger than that of $\Delta^{++}$. For $\Delta^-$, it appears that its mass has not been measured.

TABLE III
Baryon Decuplet Masses

| | Observed Masses *(MeV)* | M*(i,j) Masses |
|---|---|---|
| $\Delta^{++}$ | 1230.5 ± 0.3 (a) | *M\*(1,2)*=1230.702 |
| | 1230.9 ± 0.2 (a) | |
| | 1231.1 ± 0.2 (a) | |
| $\Delta^{+}$ | 1231.2 ± ? (a) | *M\*(1,1)*=1230.191 |
| | 1231.6 ± ? (a) | |
| | 1234.9 ± 1.4 (a) | |
| $\Delta^{0}$ | 1232.5 ± 0.3 (b) | *M\*(1,0)*=1231.484 |
| | 1233.1 ± 0.3 (a) | |
| | 1233.8 ± 0.2 (a) | |
| $\Delta^{-}$ | ? | *M\*(1,-1)*=1234.582 |
| $\Sigma^{*+}$ | 1382.8 ± 0.4 (a) | *M\*(2,1)*=1382.8 (input) |
| $\Sigma^{*0}$ | 1383.7 ± 1.0 (a) | |
| | 1384.1 ± 0.8 (b) | *M\*(2,0)*=1384.093 |
| $\Sigma^{*-}$ | 1387.2 ± 0.5 (a) | *M\*(2,-1)*=1387.191 |
| $\Xi^{*0}$ | 1531.8 ± 0.3 (a) | *M\*(3,0)*=1531.8 (input) |
| $\Xi^{*-}$ | 1535.0 ± 0.6 (a) | *M\*(3,-1)*=1534.898 |
| $\Omega^{-}$ | 1672.45 ± 0.29 (a) | *M\*(4,-1)*=1672.799 |

(a) Particle Data Group, Physics Lett. **592B** (2004)
(b) Particle Data Group, The European Physical Journal **C3** (1998)

## II. Electromagnetic Splittings

The electromagnetic mass splittings based on SU(6) model [8] are given as,

(i) $\Delta^+ - \Delta^{++} = n - p - (\Sigma^+ + \Sigma^- - 2\Sigma^0)$,
(ii) $\Delta^0 - \Delta^+ = \Sigma^{*0} - \Sigma^{*+} = n - p$,
(iii) $\Delta^- - \Delta^0 = \Sigma^{*-} - \Sigma^{*0} = \Xi^{*-} - \Xi^{*0} = n - p + (\Sigma^+ + \Sigma^- - 2\Sigma^0)$.

Using (1) &(2), it is shown readily below that the mass formulae are not only quite consistent with the above relations, but also show what the value of each difference should be in terms of three masses, namely that of *e, p,* and *n*. Following the *i* and *j* numbers given in Table I, the corresponding relations and their values are,

(i) $M^*(1,1) - M^*(1,2) = n - p - [M(2,1) + M(2,-1) - 2M(2,0)] = -e.$
(ii) $M^*(1,0) - M^*(1,1) = M^*(2,0) - M^*(2,1) = n - p.$
(iii) $M^*(1,-1) - M^*(1,0) = M^*(2,-1) - M^*(2,0) = M^*(3,-1) - M^*(3,0) =$
$n - p + [M(2,1) + M(2,-1) - 2M(2,0)] = C = 2(n-p) + e.$

When the observed masses shown in Table II and Table III are used to obtain numerical values, all of the electromagnetic splittings predicted by the SU(6) model are in excellent agreement within 0.02*(MeV)* except one case, $\Delta^+ - \Delta^{++}$ which may be attributable, as explained before to the relatively large uncertainty in the observed mass of $\Delta^+$. If the calculated masses of $M(i, j)$ and $M^*(i, j)$ are used, all of the SU(6) predictions, of course become exact as shown above.

Another relation for electromagnetic splitting based on the approximate SU(3) is known as Coleman-Glashow relation [9], $\Sigma^+ - \Sigma^- = p - n + \Xi^0 - \Xi^-$. This relation when evaluated by use of the observed masses [5] deviates by 0.27 *(MeV)*, an order of magnitude larger than that of the previous relations. This deviation can be seen readily as an expected result when we calculate the relation using *M(i, j)*,

$$\Sigma^+ - \Sigma^- = M(2,1) - M(2,-1) = -5(n-p) - 3e,$$
$$p - n + \Xi^0 - \Xi^- = p - n + M(3,0) - M(3,-1) = -5(n-p) - 5e/2.$$

The difference between the above two equations is *e*/2=0.26 *(MeV)* which is indeed close to the above mentioned observed deviation of 0.27*(MeV)*.

There are altogether 15 possible electromagnetic mass differences in baryon octet and decuplet as shown in the Table IV. Observed mass differences in the table are calculated from the data given in Table II and Table III.

TABLE IV

Electromagnetic Mass Splittings of the Octet and Decuplet Baryons

| | Observed Masses *(MeV)* | Calculated Masses *(MeV)* |
|---|---|---|
| $n - p$ | 1.29 | $M(0,0) - M(0,1) = (n - p) = 1.29$ |
| $\Sigma^0 - \Sigma^+$ | 3.10 | $M(2,0) - M(2,1) = 2(n - p) + e = 3.10$ |
| $\Sigma^- - \Sigma^+$ | 7.97 | $M(2,-1) - M(2,1) = 5(n - p) + 3e = 7.99$ |
| $\Sigma^- - \Sigma^0$ | 4.88 | $M(2,-1) - M(2,0) = 3(n - p) + 2e = 4.90$ |
| $\Xi^- - \Xi^0$ | 6.41 | $M(3,-1) - M(3,0) = 4(n - p) + 2.5e = 6.45$ |
| $\Sigma^{*0} - \Sigma^{*+}$ | 1.30 | $M^*(2,0) - M^*(2,1) = n - p = 1.29$ |
| $\Sigma^{*-} - \Sigma^{*+}$ | 4.40 | $M^*(2,-1) - M^*(2,1) = 3(n - p) + e = 4.39$ |
| $\Sigma^{*-} - \Sigma^{*0}$ | 3.10 | $M^*(2,-1) - M^*(2,0) = 2(n - p) + e = 3.10$ |
| $\Xi^{*-} - \Xi^{*0}$ | 3.20 | $M^*(3,-1) - M^*(3,0) = 2(n - p) + e = 3.10$ |
| $\Delta^0 - \Delta^+$ | 1.30 | $M^*(1,0) - M^*(1,1) = n - p = 1.29$ |
| $\Delta^0 - \Delta^{++}$ | 2.00 | $M^*(1,0) - M^*(1,2) = (n - p) - e = 0.78$ |
| $\Delta^+ - \Delta^{++}$ | 0.70 | $M^*(1,1) - M^*(1,2) = -e = -0.511$ |
| $\Delta^- - \Delta^{++}$ | ? | $M^*(1,-1) - M^*(1,2) = 3(n - p) = 3.88$ |
| $\Delta^- - \Delta^+$ | ? | $M^*(1,-1) - M^*(1,1) = 3(n - p) + e = 4.39$ |
| $\Delta^- - \Delta^0$ | ? | $M^*(1,-1) - M^*(1,0) = 2(n - p) + e = 3.10$ |

Using the above basic mass differences, other new combinatorial mass-difference relations can be easily formed such as the ones discussed above in the SU(6) predictions.

## III. Link to the Resonances

In the mass formulae $M(i, j)$ and $M^*(i, j)$, $i$ numbers go only up to 3 for the octet baryons and to 4 for the decuplet baryons. When $M(i, j)$ is plotted as a function of $i$ running beyond 14, it shows a bell-shaped curve peaking at 1842 *(MeV)* with $i=10$ and $j=-1$. For $M^*(i, j)$, it reaches the peak at 2043 *(MeV)* with $i=9$ and $j=-1$. If we group the spin ½ octet resonances according to the pattern shown in Table V and plot them together with $M(i, j)$ curve, Fig.1.shows some features that may be of some interest.

Three solid lines in the figure represent the different charge states, top one is for $j=-1$, middle one for $j=0$ and bottom line for $j=+1$. The circular points are the octet baryons, but only four points are discernible rather than the eight because the electromagnetic differences are too small to be resolved in the plot. Square points are the resonances. The resonant part of the graph, while not as good as the octet component as expected, still seems to provide, with an exception of Λ (1405) a fair agreement between the observed resonances and $M(i, j)$. For the mass formula, Λ (1405) and Λ (1116) seem to be anomaly. In the plot, it may be noteworthy that the largest observed spin ½ octet resonance at 1810 *(MeV)* seems to coincide with the near peak value of $M(i, j)$.

TABLE V
Spin ½ Baryon Octet Resonace Masses

| Observed Masses *(MeV)* | | | $M(i, j)$ Masses *(MeV)* |
|---|---|---|---|
| Λ | 1405 | ±4 | ? |
| N | 1440 | 1430 – 1470 | $M(4,-1)=1439$ |
| N | 1535 | 1520 – 1555 | $M(5,0)=1538$ |
| Λ | 1600 | 1560 – 1700 | $M(6,1)= 1623$ |
| N | 1650 | 1640 – 1680 | $M(6,0)= 1633$ |
| Σ | 1660 | 1630 – 1690 | $M(6,-1)=1644$ |
| Λ | 1670 | 1660 – 1680 | $M(6,-2)=1656$ |
| N | 1710 | 1680 – 1740 | $M(7,0)=1712$ |
| Σ | 1750 | 1730 – 1800 | $M(8,1)=1759$ |
| Λ | 1800 | 1720 – 1850 | $M(9,1)=1797$ |
| Λ | 1810 | 1750 – 1850 | $M(9,0)=1811$ |

Observed values are taken from the Baryon Summary Table in
the Particle Data Group, Physics Letters **592B** (2004), Ref. [5].

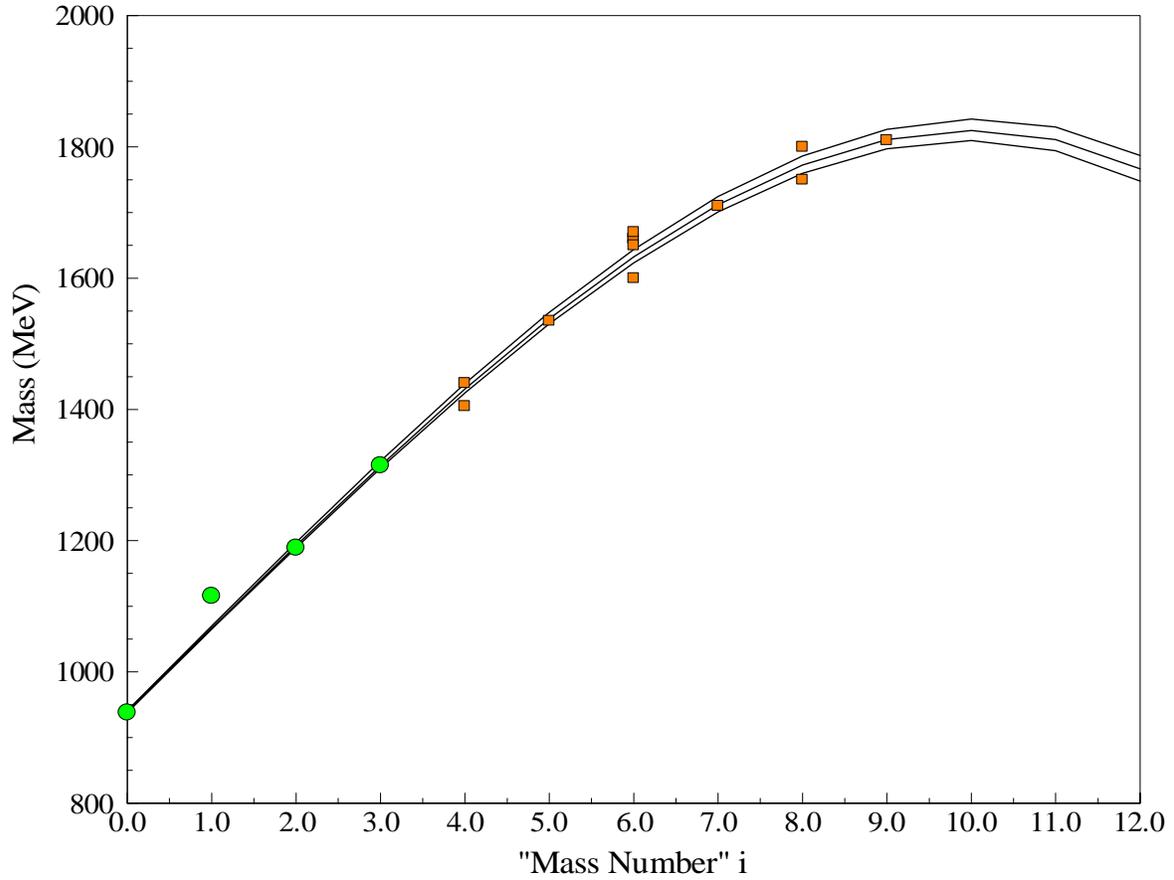

Fig. 1. Three solid lines are the plots of $M(i, j)$ with respect $i$. Top curve represents the charge state of $j=-1$, middle one $j=0$ and bottom line $j=+1$. Circular points are the octet baryons. Only four of them are discernible because the electromagnetic splittings are too close together to be resolved in this plot. Square points are the spin ½ octet resonances.

In the above plot, the spin ½ octet resonances appear to be direct extension of the octet members. In particular, the resonance at $N(1440)=M(4,-1)$ known as Roper resonance occupies the same block as $\Omega^-$ in Table I and looks like it is as much of an octet member as it is a resonance.

## IV. Summary

It is intended to show that simple formulae with one integer variable which acts like a quantum number can generate "ordinary" baryon masses with precision in most cases. The formulae also generate the same electromagnetic mass differences predicted by the SU(6) model and in addition, show what their mass differences should be. The spin ½ octet resonances appear to be the direct extension of the octet formula $M(i, j)$. Could there be a possibility that there may be some underlying connection with the quark model?